\documentclass{article}
    \usepackage{appendix}
    \usepackage{amsmath,amssymb,stmaryrd,amsthm, natbib}
    \usepackage{hyperref}
    \usepackage{framed,comment}
    \usepackage{cancel}

    \def\p{\partial}

    \DeclareMathOperator{\diff}{d}
    \newcommand{\pp}[2]{\frac{\partial #1}{\partial #2}} 
    \newcommand{\dede}[2]{\frac{\delta #1}{\delta #2}}

    \newcommand{\scp}[2]{\left<#1\,,\,#2\right>}

    \DeclareMathOperator{\sat}{sat}

    \newtheorem{theorem}{Theorem}[section]

    \newtheorem{proposition}[theorem]{Proposition}
    \newtheorem{remark}[theorem]{Remark}
     
    \usepackage[margin=2cm]{geometry}
    \usepackage{fancybox}
    \usepackage[color=yellow]{todonotes}
    \usepackage{mathrsfs}

\begin{document}
\title{Variational derivation of a moist thermal rotating shallow water model}
\author{Colin J. Cotter \and Darryl D. Holm \and Oliver D. Street}
\date{24 June 2026}

\maketitle

\begin{abstract}
    We introduce a new energy-conserving, moist shallow water model with thermal stratification and rotation. The model is derived from a variational principle, using a Lagrangian expressed in terms of enthalpy. In this model, the latent heat from phase transitions modifies the buoyancy dynamics, which in turn feeds back to alter the vertically integrated hydrodynamic motion. Finally, we generalise this moisture parameterisation to non-hydrostatic Green-Naghdi equations.
\end{abstract}

\tableofcontents

\section{Introduction}
\subsection{Background}
The coupling of physics and transport in weather and climate models is a highly active area of 
current research \citep{gross2018physics}. Prominent within this research is the effort to understand the interaction between physics parameterisations and the fluid dynamics in the ``dynamical core" of a numerical model, with an eye towards avoiding spurious artefacts that could lead to systematic errors or even
instabilities. In particular, switches due to phase transitions can trigger gridscale phenomena in the model that are more severe than those arising from the dry dynamics. Hence, the presence of phase transitions necessitates further robustness in the discretisation approach.

Moist thermal shallow water equations provide a laboratory for exploring these issues in a simplified setup that is computationally much less expensive than fully 3D models. These two dimensional models capture many of the same phenomena that govern the physics of dynamical coupling in three dimensions, such as Rossby waves and geostrophic adjustment. A model for this purpose was derived in \citet{zerroukat2015moist} by considering vertical averaging in an incompressible Boussinesq model, then using the result to investigate the behaviour of the semi-implicit semi-Lagrangian discretisation approach. The model has three moisture phases: water vapour, cloud and rain. \citet{ferguson2019assessing} used the same model to provide a challenging test case for adaptive mesh refinement, seeking to resolve sharp gradients arising from phase transitions. \citet{santos2021topography} also used it to study local grid refinement in the context of clouds and precipitation over the Andes. \citet{hartney2025exploring} also used this model to investigate the robustness of the compatible finite element approach to numerical discretisation.

Shallow water models which incorporate some combination of moisture parameterisation, thermal effects and convective adjustment have also been used to explain fundamental meteorological phenomena.
An important early work is \citet{gill1982studies}, where a linearised shallow water 
system with linearised moisture transport was considered, with a nonlinear switch in the equations between wet and dry regions. This type of modelling approach was used to describe precipitation fronts in the tropical atmosphere \citet{frierson2004large,pauluis2008precipitation}. \citet{bouchut2009fronts} extended this approach by introducing a shallow water model with coupled moisture tracer and 
full advective nonlinearity, using Riemann characteristics to demonstrate precipitation fronts as well as presenting numerical results. This model was 
described as ``moist convective", because when water vapour is condensed into cloud,
it is taken out of the shallow layer with corresponding reduction in layer depth. This process models convection into the region above the shallow layer.  \citet{lambaerts2011simplified} extended the original model by adding an upper layer so that the overall system conserves mass. The presence of this upper layer also allows for baroclinic instability. 

Another extension was to introduce an explicit temperature or buoyancy variable so that the latent heat release of condensation can be modelled directly by an increase in temperature. \citet{kurganov2020moist} proposed a model that splits the latent heat release so that part of it causes layer depth reduction modelling convection as before, and the rest is explicitly increasing the temperature. This model was obtained by 
adding appropriate source terms to the thermal shallow water model that allows for temperature variation between vertical columns \citep{ripa1993conservation}, thereby producing
the moist convective thermal shallow water equations. \citet{kurganov2021interaction} used this model to construct tropical cyclone-like vortices interacting with sea surface anomalies, and \citet{rostami2022genesis} showed that it can exhibit Madden-Julien type oscillations. This modelling approach has been extended to sophisticated multilayer shallow water models in the Potsdam Climate Institute intermediate complexity model Aeolus \citep{rostami2025aeolus}. 

Shallow water models with convection parameterisations have also been proposed for testing data assimilation schemes, due to their increased nonlinearity and unpredictability. \citet{wursch2014simple} proposed a model with a term that adds
divergence in the velocity equation in regions satisfying a condition for convection,
causing localised peaks in the layer depth representing convective columns. \citet{kent2017dynamics} studied the hyperbolic structure of the 1D inviscid version
of this model, and \citet{cantarello2022idealized,bokhove2022idealized} added vertical structure by extending
to a two layer model with a rigid lid (referred to as a $1+1/2$ layer model).

In a parallel direction, there have also been recent efforts to understand and improve the energetic and thermodynamic consistency of atmosphere models \citep{gassmann2013global,gassmann2021inherent,lauritzen2022reconciling,eldred2022thermodynamically,guba2024energy,dubos2024thermodynamic,bowen2022consistent,ricardo2024thermodynamic,thuburn2025potential}. Similar efforts have also been pursued for ocean models
\citep{young2010dynamic,tailleux2013available,tailleux2024simple} to account for the presence of both salinity and entropy in the  equation of state for the seawater Boussinesq system.
A foundational variational framework for guaranteeing energetic and 
thermodynamic consistency in fluid models was established in \citet{gay2017lagrangian,gay2018dirac,gay2020dirac}. Their variational framework was applied to include moisture in atmosphere models in \citet{gay2019variational,eldred2021thermodynamically}.

\subsection{The present work}
In this article, we consider a new moist thermal shallow water model, derived by using the variational approach introduced in \citet{gay2019variational}. Notably, this variational derivation guarantees the conservation of energy as well as a consistent Kelvin circulation law, thereby including the discussion of energetic consistency as part of the examination of the coupling of physics and transport in shallow water models.

We shall consider a three species model where the contributions to the layer depth $D$ are divided into dry air $D_d$, water vapour $D_v$, and cloud water $D_c$, such that $D=D_d+D_v+D_c$. Following the notation of \cite{gay2019variational}, we shall use the index
$k \in \{d,v,c\}=:\mathcal{K}$ to sum over these components. As in the traditional shallow water equations, the total layer depth $D$ is advected as a density. The separate contributions to the layer depth evolve over time as
\begin{equation}\label{eqn:depth_SW}
    \frac{\partial D_d}{\partial t} + \nabla\cdot(uD_d) = 0 \,,\quad
    \frac{\partial D_v}{\partial t} + \nabla\cdot(uD_v) = j \,,\quad
    \frac{\partial D_c}{\partial t} + \nabla\cdot(uD_c) = -j \,,
\end{equation}
where $j$ is set to ensure that the following complementarity condition, for mixing ratios $q_k=D_k/D$, $\forall k\in \mathcal{K}$, is satisfied
\begin{align}
    q_v - q_{\sat} &\leq 0
    \,,\\
    q_c &\geq 0
    \,,\\
    (q_v - q_{\sat})q_c &= 0
    \,.
\end{align}
Here $q_{\sat} = q_{\sat}(b,D)$ is the saturation concentration function, specifying the maximum fraction of water vapour at a given buoyancy and total layer depth. 
The above complementarity condition can be expressed as
\begin{equation}
    q_v = {\rm min}(q_v+q_c,q_{\sat}) \,.
\end{equation}
This means that either $q_v < q_{\sat}$ and $q_c=0$ (the air parcel is subsaturated and there is no cloud in it), or $q_c>0$ and $q_v=q_{\sat}$ (the air parcel is saturated and any excess moisture has condensed into cloud). Consequently, any excess water vapour
is instantaneously condensed into cloud. Conversely, if the air parcel is subsaturated
then any cloud water is instantaneously evaporated (transformed to vapour) as far is allowed by
$q_{\sat}$. The functional dependence in $q_{\sat}$ on buoyancy ($b$) and depth ($D$) is a modelling choice
which may be derived from a chosen saturation value in the three dimensional parent
model by considering column averaging. For an example, see Appendix A3 of \citet{zerroukat2015moist}, reinterpreted for a buoyancy variable in \citet{hartney2025exploring}.

The equation for the evolution of 2D velocity $\boldsymbol{u} = (u,v)$ takes the same form as in conventional theories of thermal shallow water dynamics; namely,
\begin{equation}\label{eqn:u_SW}
    \p_t\boldsymbol{u} + \boldsymbol{u}\cdot\nabla\boldsymbol{u} + f\boldsymbol{u}^\perp 
    = gb\nabla (D-H) + \frac12 gD\nabla b
    = \frac{g}{2}\nabla(bD) +\frac{g}{2}  b\nabla (D-2H) 
    \,,
\end{equation}
where $f$ is the Coriolis parameter, $g$ is a gravitational constant, and $\boldsymbol{u}^\perp = (-v,u)$. Here, the buoyancy variable $b$ evolves according to
\begin{equation}\label{eqn:buoyancy_SW}
    \partial_t b + u\cdot\nabla b 
    =\frac{2jL}{gD\left(-D+2H - \frac{2B}{A}\right)} \,,
\end{equation}
where $L$ is a latent heat constant, $z=-H(\boldsymbol{x})$ is the bottom of the domain, and $A,B \in \mathbb{R}$ are thermodynamic constants with $A\neq 0$, depending on the reference profile and discussed in Section \ref{sec:averaged_thermodynamics}. The complete system of equations is thus given by equations \eqref{eqn:depth_SW}, \eqref{eqn:u_SW}, and \eqref{eqn:buoyancy_SW}.

Due to the method of derivation used (discussed in subsequent sections), the model introduced here has several conservation laws. In particular, the latent heat release term in \eqref{eqn:buoyancy_SW} ensures that the system conserves total energy.

\subsubsection*{Plan of the paper.}
The rest of this article is structured as follows.
\begin{itemize}
\item Section \ref{sec:averaged_thermodynamics} derives the modified Lagrangian for thermal shallow water dynamics. The Lagrangian used here is represented in terms of enthalpy rather than internal energy. 
\item Section \ref{sec: DeriveMoistSW} derives the moist shallow water model obtained by following the approach of \cite{gay2019variational} which is based on thermodynamic enthalpy and combines free variations of enthalpy in mass and entropy with constrained phenomenological dissipative variations augmented by a term representing the latent heat release.
\item The moist shallow water model derived in section \ref{sec: DeriveMoistSW} is extended in section \ref{sec: DeriveMoistGN} by following the same enthalpy method in deriving the moist Green-Naghdi model, which includes nonhydrostatic effects in shallow water dynamics. 
\item Section \ref{sec: summary} provides a summary of these results and presents ideas for outlook, open problems and future challenges of the geometric-thermodynamic approach to fluid dynamics with phase changes.
The summary section \ref{sec: summary} is followed by three appendices which provide details of the calculations described in the body of the text. 
\item Appendix \ref{app:formulation} describes the moisture parameterisation in the modified Hamilton's principle. 
\item Appendix \ref{app:Lagrangian} discusses the Lagrangian for thermal Euler-Boussinesq dynamics, expressed in terms of enthalpy. Moreover, using the Language developed in Appendix \ref{app:formulation}, it demonstrates that this Lagrangian indeed corresponds to the Euler-Boussinesq equations through Hamilton's Principle. 
\item Appendix \ref{app: C-EnergyConserved} proves conservation of energy for the moist shallow water model. This is a result of applying the standard Euler--Poincar\'e (EP) approach to Hamilton's variational principle for deriving 2D thermal rotating shallow water equations. 
\end{itemize}

\section{Approximating the Boussinesq model}\label{sec:averaged_thermodynamics}

We seek to formulate a Lagrangian which is suitable for vertical integration, to derive shallow water models, beginning from the above Lagrangian for the Euler-Boussinesq equation and approximating with vertically averaged variables. The Lagrangian appears in the appendices (see equation \eqref{eqn:EB_Lagrangian}) and is repeated here as
\begin{equation}\label{eqn:EB_Lagrangian_repeated}
    \ell_{EB}(\boldsymbol{v},\rho,\eta;p') = \int \frac{\rho|\boldsymbol{v}|^2}{2} + \rho \boldsymbol{v}\cdot\boldsymbol{R} - \rho\phi - \rho h(p_0,\eta) + p'\left( 1-\frac{\rho}{\rho_0} \right) + p_0\,d^3x \,,
\end{equation}
In this Lagrangian, $\boldsymbol{v}$ is the three dimensional velocity field, $\rho$ is density with constant reference density $\rho_0$, $\eta$ is (scalar) entropy, $\phi$ is the geopotential, $p_0$ is the hydrostatic pressure, and $p'$ is a Lagrange multiplier enforcing incompressibility. For a full discussion of this Lagrangian, and a demonstration that it corresponds to the Boussinesq model, see Appendix \ref{app:Lagrangian}.

For vertically averaged models, we will denote the $x-y$ domain by $\Omega$, the bottom topography by $z=-H(x,y)$, the free upper boundary by $z=\zeta(x,y,t)$, and the depth by $D=\zeta+H$.  In order to properly isolate the $z$-dependence, we must substitute in a background pressure profile. As in \cite{eldred2021thermodynamically}, we consider the case where we have a stratified background profile in hydrostatic balance. That is
\begin{equation}\label{eqn:geostrophic_balance}
    \nabla p_0 (x) = -\rho_0\nabla\phi(x) \,,
\end{equation}
and, when the geopotential is taken to be $\phi=gz$, this gives
\begin{equation}\label{eqn:geostrophic_balance_integrated}
    p_0 = -\rho_0gz + p_{00} \,,
\end{equation}
where $p_{00}$ is some constant reference value of the pressure. For geostrophic balance, the fluid is assumed to be in its rest state with the boundary condition $p_0(z=0) = p_{00}$. To make sense of this background profile over $z\in[-H,\zeta]$, we must extend this definition by expanding around $z=0$ as
\begin{equation}
	p_0(z) = p_0(z=0) + z\frac{\partial p_0}{\partial z}\bigg|_{z=0} + \dots = p_{00} - \rho_0gz   \,.
\end{equation}
Thus, for small perturbations of the surface, the definition \eqref{eqn:geostrophic_balance_integrated} is reasonable for $z\in[-H,\zeta]$. Continuing, the Lagrangian also contains nontrivial $z$ dependence in its enthalpy term. To handle this we expand its $p_0$ dependence similarly (whilst holding $\eta$ fixed) to find that
\begin{equation}
\begin{aligned}
	h(p_0(z),\eta) &= h(p_0(z=0),\eta) + z \frac{\partial h(\tilde p,\eta)}{\partial \tilde p}\bigg|_{\tilde p = p_0}\frac{\partial p_0}{\partial z}\bigg|_{z=0} 
	\\
	&= h(p_{00},\eta) - z\rho_0 g\frac{1}{\rho(p_0,\eta)}  = h(p_{00},\eta) - gz
	\,.
\end{aligned}
\end{equation}
In the final line, we have used that $\rho(p_0,\eta) = \rho_0$ in the Boussinesq approximation and the thermodynamic relationship $1/\rho = \partial h / \partial p$.

It remains to identify the desired $z$ dependence in the remaining enthalpy term. Using the ideal gas equation $e=c_vT$ and thermodynamic First Law $T = \partial e / \partial \eta $, we see that
\begin{equation}
    e = \Phi(\alpha) \exp{\frac{\eta}{c_v}} \,,
\end{equation}
where $\alpha=1/\rho$ is the specific volume and $\Phi(\alpha)$ is some arbitrary function. Since $\rho e(\rho,\eta) = \rho h(p,\eta) - p$ and $p=-\partial e / \partial \alpha$, we have that $h(p,\eta) = (\Phi(\alpha) - \alpha \Phi'(\alpha))\exp{\frac{\eta}{c_v}}$. Denoting by $\eta_0$ the entropy attributed to the hydrostatic rest configuration, and thus introducing a decomposition of total entropy $\eta=\eta_0+\eta'$, we note that enthalpy at $p_{00}$ takes the form
\begin{equation}
    h(p_{00},\eta) = h(p_{00},\eta_0)\exp \frac{\eta'}{c_v} = \Psi(z)\exp \frac{\eta'}{c_v}\,.
\end{equation}
The condition that our background reference configuration is in hydrostatic balance places no restriction on its thermodynamic state. If we assume that it is an ideal gas, and hence that $h=c_pT$, we see that if our reference profile is such that it has a constant lapse rate (such as it is in the troposphere), $\Gamma$ then $\Psi$ is linear in $z$. Indeed, we may write $\Psi(z) = Az+B$ where $A = c_p\Gamma$ and $B$ is the value of the enthalpy in the reference profile at $z=0$. The sign and magnitude of $A$ correspond to the direction and strength of the assumed thermodynamic profile.  

In the shallow water case, the volume form $d^3x$ becomes the layer thickness weighted area form $D\,d^2x$, where $D = \int dz$. As such, we write the enthalpy in terms of the (scaled) shallow water entropy density $s = D\eta'/(2c_v)$ as
\begin{equation}
    h(p_{00},s) = ( A z + B) \exp \frac{2s}{D} \,.
\end{equation}
For shallow water dynamics, we will be assuming that $s$ is depth independent, which serves to limit the thermodynamic activity which is permitted in the vertical direction whilst permitting horizontal moisture features. Following the expansion of the potential energy of the fluid, also illustrated for the Boussinesq approximation in equation \eqref{eqn:pe_expansion} of Appendix \ref{app:Lagrangian}, we see that our potential energy is
\begin{equation}
    \rho gz + \rho e(\rho,\eta) \approx \rho_0 gz + \rho_0 h(p_{00},\eta) - p_{00} = \rho_0 g z + \rho_0(Az+B)\exp\frac{2s}{D} - p_{00} \,,
\end{equation}
where we have evaluated $\rho = \rho_0$. Making further the `traditional approximation' on the Coriolis force, and denoting the three dimensional velocity field by $v = (u,w)$, where $u$ is the (assumed $z$-independent) horizontal velocity field and $w$ is the vertical velocity, the Lagrangian \eqref{eqn:EB_Lagrangian_repeated} becomes
\begin{equation}\label{eqn:EB_Lagrangian_approximated}
    \ell_{approx}(\boldsymbol{u},w,s,D) = \int_{\Omega}\int_{-H}^\zeta \frac{1}{2}\left(|\boldsymbol{u}|^2 + w^2 \right) + \boldsymbol{u}\cdot\boldsymbol{R} -gz - (A z + B) \exp \frac{2s}{D} + \frac{p_{00}}{\rho_0}\,d^3x \,.
\end{equation}
The final term does not contribute to the evolution equations since it is a constant. For the shallow water equations, we make the hydrostatic approximation by dropping the $w$-dependence of the kinetic energy. To deduce the Green-Naghdi type model, we obtain an expression for $w$ by vertically integrating the incompressibility constraint.

Following the definition in \cite{eldred2021thermodynamically}, the form of the potential energy allows us to define a buoyancy variable, 
\begin{equation}
    b = -\frac1g\frac{\partial}{\partial z}\left((Az+B)\exp\frac{2s}{D} + gz\right) = -\frac{A}{g}\exp \frac{2s}{D} - 1 \,.
\end{equation}
Here, we remark that the condition imposed on the reference configuration having a constant lapse rate, and hence enthalpy being linear in depth, is precisely the same condition as the buoyancy variable being depth independent.

\section{Derivation of the moist shallow water model}\label{sec: DeriveMoistSW}

We derive the model by applying the variational methodology summarized in the Appendix \ref{app:formulation} to a suitable Lagrangian. To minimise the geometric notation required to illustrate the equations and their derivation, we will be using a vector calculus friendly notation described in Remark \ref{rmk:notation}. As explained in Section \ref{sec:averaged_thermodynamics}, we perform a vertical average of the Lagrangian given in equation \eqref{eqn:EB_Lagrangian_approximated}. We consider a flow domain with $(x,y)\in\Omega \subseteq \mathbb{R}^2$. The bottom topography is denoted by $z=-H(x,y)$; the free upper boundary by $z=\zeta(x,y,t)$; and the depth by the sum $D=\zeta+H$. In this flow domain, we assume that the velocity variables are independent of vertical coordinate; thereby admitting columnar motion and enabling vertical averaging. The Lagrangian for thermal shallow water dynamics is derived from equation \eqref{eqn:EB_Lagrangian_approximated} in Section \ref{sec:averaged_thermodynamics} by making these depth-independence assumptions, applying the constraint that density is constant, and making a hydrostatic approximation by neglecting the dependence of the kinetic energy on vertical velocity. When applying the constraint that density is constant, we remove the Lagrange multiplier constraint and substitute the result, $\rho=\rho_0$, back into the Lagrangian. The Lagrangian is derived by performing the vertical average as
\begin{equation}
\begin{aligned}
    \ell_{SW}(\boldsymbol{u},D,s) &= \int\left[ \int_{-H}^\zeta \left(\frac{|\boldsymbol{u}|^2}{2} + \boldsymbol{u}\cdot\boldsymbol{R} -gz - (A z+B)\exp \frac{2s}{D} + \frac{p_{00}}{\rho_0} \right)dz\right]d^2x 
    \\
    &= \int \frac{D|\boldsymbol{u}|^2}{2} + D\boldsymbol{u}\cdot\boldsymbol{R} - \frac{1}{2}\left( A \exp \frac{2s}{D} + g \right)D(D-2H) - DB\exp\frac{2s}{D} + D\frac{p_{00}}{\rho_0} \,d^2x \,,
\end{aligned}
\end{equation}
where the exponential form of the enthalpy, with thermodynamic constants $A,B \in \mathbb{R}$ such that $A\neq 0$, is discussed in Section \ref{sec:averaged_thermodynamics}, and each variable in the Lagrangian $\ell_{SW}$ depends only on $x,y \in \Omega$. The final term in the Lagrangian does not influence either the dynamics or the thermodynamics since both the reference pressure and density, $p_{00}$ and $\rho_0$ respectively, are constant as described in Section \ref{sec:averaged_thermodynamics}. The \emph{moist} shallow water model corresponds to this Lagrangian, augmented by adding the term $LD_v$ which represents the latent heat release according to the Landau paradigm, as discussed in \cite{landau1959lifshitz,Umantsev2023},%
\footnote{In addition to dependence on entropy ($s$), pressure ($p$), and composition ($D_k$) for the enthalpy equilibrium, the Landau paradigm adds dependence on a set of continuum \emph{order parameters} to the free energy which are associated with phase changes. According to the laws of thermodynamics, an equilibrium state of an open system with given p and s corresponds to the minimum of the enthalpy. Hence, the equilibrium will be found among the critical points of the enthalpy as a function of the order parameter, which is the latent heat $L$ in the present case.}
yielding
\begin{equation}
    \ell_{MSW}(\boldsymbol{u},\{D_k\},s) = \int \frac{D|\boldsymbol{u}|^2}{2} + D\boldsymbol{u}\cdot\boldsymbol{R} - \frac{1}{2}\left( A \exp \frac{2s}{D} + g \right)D(D-2H) - DB\exp\frac{2s}{D} + LD_v \,d^2x \,.
\end{equation}
Notice that the addition of the latent heat term does not violate the ideal gas assumption used to derive the form of the enthalpy, since this augmentation can be interpreted as replacing dry static energy with moist static energy.
To deduce our model equations, we compute the variational derivatives according to their definitions, e.g., 
\begin{equation}\label{eqn:var_derivs_def_2D}
\begin{aligned}
    \delta \ell(u(\boldsymbol{x},t)) &:= \frac{\p}{\p\epsilon}\bigg|_{\epsilon=0} \ell(u(\boldsymbol{x},t,\epsilon)) = \scp{\frac{\delta\ell}{\delta u}} {\frac{\p u}{\p\epsilon}\bigg|_{\epsilon=0}} =: \scp{\frac{\p\ell}{\p u}}{ \delta u } := \int\sum_{j=1}^2\frac{\p\ell}{\p u^j} \,\delta u^j d^2x = \int \frac{\delta\ell}{\delta\boldsymbol{u}}\cdot\delta\boldsymbol{u}\,d^2x
    \,,\\
    \delta\ell(s(\boldsymbol{x},t)) &= \left\langle \frac{\delta\ell}{\delta s} , \delta s \, d^2x \right\rangle :=  \int \frac{\delta\ell}{\delta s}\delta s \,d^2x
    \,,\quad\hbox{and}\quad
    \delta\ell(D_k(\boldsymbol{x},t)) = \int \frac{\delta\ell}{\delta D_k}\delta D_k \,d^2x
    \,,\quad\hbox{for each $k$.}
\end{aligned}
\end{equation}
Note that, according to these definitions, the variational derivatives do not carry bases with them and are either scalar or vector valued. Vector fields are denoted in bold font when expressed in terms of their basis, i.e. $v=\boldsymbol{v}\cdot\nabla$. The variational derivatives are
\begin{align}
    \frac{\delta \ell_{MSW}}{\delta \boldsymbol{u}} &= D(\boldsymbol{u}+\boldsymbol{R})
    \,,\\
    \frac{\delta\ell_{MSW}}{\delta D_d} &= \frac{|\boldsymbol{u}|^2}{2} + \boldsymbol{u}\cdot\boldsymbol{R} - (D-H)g + A\exp \frac{2s}{D}\left( H-D + s\frac{D-2H}{D} \right) 
    + {B\exp\frac{2s}{D} \Big(\frac{2 s}{D} - 1 \Big)} + p_0,
    \\
    &=: \varpi
    \,,\\
    \frac{\delta\ell_{MSW}}{\delta D_c} &= \varpi
    \,,\\
    \frac{\delta\ell_{MSW}}{\delta D_v} &= \varpi + L
    \,,\\
    \frac{\delta \ell_{MSW}}{\delta s} &= -(D-2H)A\exp\frac{2s}{D} - 2B\exp\frac{2s}{D} = -\exp\frac{2s}{D}\big( A(D-2H) + 2B \big),
    \\
    &=: \big( A(D-2H) + 2B \big)\frac{g(b+1)}{A}
    \,.
\end{align}
In the final line, we have defined the relation between buoyancy $b$ and entropy density $s$. Namely, we have defined
\begin{equation}
    gb = -A\exp \frac{2s}{D} - g \,,\quad\hbox{and}\quad s = \frac{D}{2}\log\frac{-g(b+1)}{A} \,.
\end{equation}
By construction, equations \eqref{eqn:depth_SW} arise by choosing $j_d = 0$, $j_v=-j_c=j$. We then assemble the variations into equations \eqref{eq:eliminated EP} and \eqref{eq:entropy}, in Appendix \ref{app:formulation}, to deduce the system. From equation \eqref{eq:eliminated EP}, the Euler-Poincar\'e momentum equation is 

\begin{equation}
\begin{aligned}
    \p_t\boldsymbol{u} + \boldsymbol{u}\cdot\nabla\boldsymbol{u} + \nabla\frac{|\boldsymbol{u}|^2}{2} + (\nabla^\perp\cdot\boldsymbol{R})\boldsymbol{u}^\perp  + \nabla(\boldsymbol{u}\cdot\boldsymbol{R})
    &= \frac{1}{D}\left( (D_d + D_c + D_v)\nabla\varpi + D_v\nabla L + s\nabla\frac{\delta\ell_{MSW}}{\delta s} \right)
    \\
    = \nabla\left( \frac{|\boldsymbol{u}|^2}{2} + \boldsymbol{u}\cdot\boldsymbol{R}  \right) 
    &+ \nabla\left( A\exp \frac{2s}{D}\left( H-D + s\frac{D-2H}{D} \right) 
    + B\exp\frac{2s}{D}\Big(\frac{2 s}{D} - 1\Big) \right)
\\
-g\nabla (D-H)&+ \frac{D_v\nabla L}{D}  - \frac{s}{D} \nabla\left((D-2H)A\exp\frac{2s}{D} + 2B\exp\frac{2s}{D}\right) 
    \,.
\end{aligned}
\end{equation}
Since the latent heat parameter $L$ is taken to be a constant, the right hand side of the momentum equation can be manipulated into
\begin{equation}
    \p_t\boldsymbol{u} + \boldsymbol{u}\cdot\nabla\boldsymbol{u} + \nabla\frac{|\boldsymbol{u}|^2}{2}  + (\nabla^\perp\cdot\boldsymbol{R})\boldsymbol{u}^\perp  + \nabla(\boldsymbol{u}\cdot\boldsymbol{R})
 = \nabla\left(\frac{|\boldsymbol{u}|^2}{2} + \boldsymbol{u}\cdot\boldsymbol{R}\right) - \left(A\exp\frac{2s}{D}+g\right)\nabla(D-H) - \frac{D}{2}\nabla\left( A\exp\frac{2s}{D}\right) \,.
\end{equation} 
Cancelling terms, and recalling our definition of the buoyancy variable, $gb = -A\exp \frac{2s}{D} - g$, this can be rewritten as
\begin{equation}
    \p_t\boldsymbol{u} + \boldsymbol{u}\cdot\nabla\boldsymbol{u} + f\boldsymbol{u}^\perp  = gb\nabla(D-H) + \frac{g}{2}D\nabla b  \,,
\label{eqn: motion}\end{equation}
where $f = \nabla^\perp\cdot\boldsymbol{R}$ and $\nabla^\perp=(\partial_y, -\partial_x)$.
This equation in turn yields the Kelvin circulation dynamics,
\begin{equation}
\label{eq:Kelvin TSW form}
\frac{d}{dt}
\oint_{c(u)}( \boldsymbol{u} + \boldsymbol{R} )\cdot d\boldsymbol{x}   = \frac{g}{2} \oint_{c(u)}  b\nabla (D-2H)  \cdot d\boldsymbol{x}
\,.
\end{equation}
We notice that the latent heat term is not modifying the forcing terms typically seen in thermal shallow water dynamics. Indeed, the latent heat term does not appear in the momentum equation explicitly. However, the latent heat does influence the motion via its effect on the evolution of the buoyancy 
in equation \eqref{eqn: b-dynamics}, below.

Instead of a forcing term in the motion equation, we obtain a source term on the right hand side of the buoyancy equation. Indeed, the buoyancy equation follows from the entropy equation \eqref{eq:entropy} in Appendix \ref{app:formulation}, which we repeat here,
\begin{equation}
\label{eq:entropy repeated}
\pp{s}{t} + \nabla\cdot\left(s\mathbf{u}\right) = -\frac{1}{\dede{\ell}{s}}\sum_k j_k\dede{\ell}{\rho_k}.
\end{equation}
Inserting the variational derivatives into \eqref{eq:entropy repeated}, we have
\begin{equation}
    \p_t\left( \frac{D}{2}\log \frac{-g(b+1)}{A} \right) 
    + \nabla\cdot\left(\left( \frac{D}{2}\log \frac{-g(b+1)}{A} \right)\mathbf{u}\right)
    = \frac{1}{\big( -D+2\left(H - \frac{B}{A}\right) \big)g(b+1)}jL
\,.
\label{almost-there}
\end{equation}
The left hand side of \eqref{almost-there} is equal to
\begin{equation}
    \frac{D}{2(b+1)}(\p_t+\mathbf{u}\cdot\nabla)b \,,
\end{equation}
and so we have the equation for buoyancy $b$:
\begin{equation}
    \p_tb+\mathbf{u}\cdot\nabla b = \frac{2(b+1)}{D}\frac{jL}{\big(  -D+2\left(H - \frac{B}{A}\right)\big)g(b+1)} 
    = \frac{2jL}{gD\left(-D+2H - \frac{2B}{A}\right)} \,.
\label{eqn: b-dynamics}
\end{equation}

Moreover, these equations also possess a clear formulation of potential-vorticity dynamics. As explained in the general case in Appendix \ref{app:formulation}, taking a `curl' operator (expressed in terms of $\nabla^\perp$ in two dimensional vector calculus) through the Kelvin-Noether form of the momentum equation gives the potential vorticity form of the equations. From equation \eqref{eqn:general_PV}, combined with the calculations above, we see that the evolution of potential vorticity (PV, denoted $q$) is driven by the misalignment of gradients,
\begin{equation}
	\p_t q + \boldsymbol{u}\cdot\nabla q 
    = \frac{g}{2D} \nabla^\perp b\cdot \nabla (D-2H) 
     = \frac{g}{2D} J\big(b,(D-2H)\big) 
    \,,
\end{equation}
where $J(\cdot,\cdot)$ is the Jacobian and 
\begin{equation}
	q:= \frac{1}{D}\left( \nabla^\perp\cdot\boldsymbol{u} + f \right) \,. 
\end{equation}
As demonstrated for the general case in Appendix \ref{app: C-EnergyConserved}, the model also conserves an energy. 
\begin{proposition}
    The moist shallow water model, given by equations \eqref{eqn:depth_SW}, \eqref{eqn:u_SW}, and \eqref{eqn:buoyancy_SW}, conserves the total energy
    \begin{equation}
        E_{MSW} = \int_\Omega \frac{D|u|^2}{2} + \frac12 gbD(D-2H) + DB\exp\frac{2s}{D} - D\frac{p_{00}}{\rho_0} - LD_v\,d^2x \,.
    \end{equation}
\end{proposition}

\section{Derivation of the moist Green-Naghdi model}\label{sec: DeriveMoistGN}

To include non-hydrostatic pressure effects, a moist Green-Naghdi model may be found by following \cite{HS2025_GN} and making similar approximations to the Lagrangian as were made in the previous section, whilst retaining the dependence of the kinetic energy on the vertical velocity. For this purpose, we use the incompressibility condition to write the vertical velocity in terms of the horizontal one. As in the previous section, we will be assuming that the horizontal velocity $u$ is dependent only on the horizontal coordinates. The incompressibility condition, in terms of the two dimensional gradient $(\nabla)$ is
\begin{equation}
    \p_zw = - \nabla\cdot \boldsymbol{u} \,.
\end{equation}
Before integrating this vertically, we first remark that the kinematic boundary condition at the upper surface of our domain is
\begin{equation}\label{eqn:kinematic_upper_boundary_condition}
    \p_t \eta + \boldsymbol{u}|_{z=\zeta}\cdot\nabla \zeta = w|_{z=\zeta} \,.
\end{equation}
The boundary condition on the bottom is
\begin{equation}\label{eqn:lower_boundary_condition}
    \boldsymbol{u}|_{z=-H}\cdot\nabla H = -w|_{z=-H} \,.
\end{equation}
Upon integrating the incompressibility constraint from the bottom of the domain to $z$, one finds
\begin{equation}
\begin{aligned}
    w &= - \int_{-H}^z \nabla\cdot\boldsymbol{u} dz' + w|_{z=-H} = - \nabla\cdot\int_{-H}^z \boldsymbol{u} \,dz' 
    \\
    &= -\nabla\cdot(\boldsymbol{u}(z+H)) = -(z+H)\nabla\cdot\boldsymbol{u} - \boldsymbol{u}\cdot\nabla H \,.
\end{aligned}
\end{equation}
The Lagrangian for moist Green-Naghdi then emerges as that of moist shallow water, plus the vertical kinetic energy restored as
\begin{equation}
\begin{aligned}
    \ell_{MGN}(u,s,\{D_k\}) &= \int\left[ \int_{-H}^\zeta \frac{|\boldsymbol{u}|^2}{2} + \frac{w^2}{2} + \boldsymbol{u}\cdot\boldsymbol{R} - gz - (A z+B)\exp \frac{2s}{D} + \frac{p_{00}}{\rho_0} \,dz\right] + LD_v d^2x
    \\
    &= \int\Bigg[ \int_{-H}^\zeta \frac{|\boldsymbol{u}|^2}{2} + \frac{1}{2}\big((z+H)\nabla\cdot\boldsymbol{u} + \boldsymbol{u}\cdot\nabla H\big)^2 + \boldsymbol{u}\cdot\boldsymbol{R} 
    \\
    &\hspace{110pt} - gz - A z\exp \frac{2s}{D} - B\exp\frac{2s}{D} + \frac{p_{00}}{\rho_0} \,dz\Bigg] + LD_v d^2x
    \\
    &= \int D\frac{|\boldsymbol{u}|^2}{2} +D\boldsymbol{u}\cdot\boldsymbol{R} + \frac16 D^3(\nabla\cdot\boldsymbol{u})^2 + \frac12D^2 (\nabla\cdot\boldsymbol{u})\left(\boldsymbol{u}\cdot\nabla H\right)  + \frac12D\left(\boldsymbol{u}\cdot\nabla H\right)^2 \\
    &\qquad - \frac{1}{2}\left( A \exp \frac{2s}{D} + g \right)D(D-2H) - DB\exp\frac{2s}{D} + D\frac{p_{00}}{\rho_0} + LD_v \,d^2x
\end{aligned}
\end{equation}
As before, we compute the variational derivatives. The variation in velocity gives
\begin{equation}
\begin{aligned}
    \frac{\delta\ell_{MGN}}{\delta u} &= D(\boldsymbol{u} + \boldsymbol{R})-\frac{1}{3}\nabla\left( D^3(\nabla\cdot\boldsymbol{u}) \right) + \frac12 D^2(\nabla\cdot\boldsymbol{u})\nabla H - \frac{1}{2}\nabla\left( D^2(\boldsymbol{u}\cdot\nabla H) \right) + D(\boldsymbol{u}\cdot\nabla H) \nabla H 
    \\
    &= D(\boldsymbol{u} + \boldsymbol{R}) - \nabla(D^2F) + DG\nabla H \,,
\end{aligned}
\end{equation}
where the quantities $F$ and $G$ are defined by
\begin{equation}
    F = \frac13 D \nabla\cdot\boldsymbol{u} + \frac12 \boldsymbol{u}\cdot\nabla H  
    \,,\quad\hbox{and}\quad
    G = \frac12 D \nabla\cdot \boldsymbol{u} + \boldsymbol{u}\cdot\nabla H  \,.
\end{equation}
Varying with respect to the entropy and the contributions to the layer depth gives
\begin{align}
\begin{split}
    \frac{\delta\ell_{MGN}}{\delta D_d} &= \frac{|\boldsymbol{u}|^2}{2} + \boldsymbol{u}\cdot\boldsymbol{R} + \frac12D^2(\nabla\cdot\boldsymbol{u})^2 + D(\nabla\cdot\boldsymbol{u})(\boldsymbol{u}\cdot\nabla H) + \frac12(\boldsymbol{u}\cdot\nabla H)^2
    \\
    &\qquad - (D-H)g + A\exp \frac{2s}{D}\left( H-D + s\frac{D-2H}{D} \right) + \frac{2B s}{D}\exp\frac{2s}{D} - B\exp\frac{2s}{D} + \frac{p_{00}}{\rho_0}
\end{split}
    \\
    &= \frac{\delta\ell_{MGN}}{\delta D_c}
    \,,\\
    \frac{\delta\ell_{MGN}}{\delta D_v} &= \frac{\delta\ell_{MGN}}{\delta D_d} + L
    \,,\\
    \frac{\delta \ell_{MGN}}{\delta s} &= -(D-2H)A\exp\frac{2s}{D} - 2B\exp\frac{2s}{D} = -\exp\frac{2s}{D}\big( A(D-2H) + 2B \big)
    \\
    &=: \big( A(D-2H) + 2B \big)\frac{g(b+1)}{A}
    \,.
\end{align}

These can be assembled into the Euler-Poincar\'e equation \eqref{eq:eliminated EP} to give the momentum equation, though the calculation is lengthier than for the hydrostatic shallow water case. This calculation is similar to that which is known for the `dry' Green-Naghdi model, see e.g. \cite{HS2025_GN}, with some small modifications. Following the calculation in the previous section for the moist shallow water model with appropriate modifications yields the Green-Naghdi motion equation,
\begin{equation}
\begin{aligned}
    \left( \p_t + \mathcal{L}_u \right)\frac{\delta\ell_{MGN}}{\delta u} &= D\bigg( \p_t\boldsymbol{u} + \boldsymbol{u}\cdot\nabla\boldsymbol{u} + u_j\nabla u^j + (\nabla^\perp\cdot\boldsymbol{R})\boldsymbol{u}^\perp + \nabla(\boldsymbol{u}\cdot\boldsymbol{R}) - \frac{1}{D}\nabla\left( D^2\frac{dF}{dt} \right) + \frac{dG}{dt}\nabla H
    \\
    &\qquad + \nabla\left( \frac12(D\nabla\cdot\boldsymbol{u})^2 + (D\nabla\cdot\boldsymbol{u})\left( \boldsymbol{u}\cdot\nabla H \right) + \frac12 \left( \boldsymbol{u}\cdot\nabla H \right)^2  \right) \bigg)\cdot d\boldsymbol{x} \,.
\end{aligned}
\end{equation}
The right hand side of the equation \eqref{eq:eliminated EP} is given by
\begin{equation}
\begin{aligned}
    \sum_k \diff\dede{\ell}{D_k}
\otimes D_k + \diff\dede{\ell}{s}\otimes s &= D\nabla\left( \frac{|\boldsymbol{u}|^2}{2} + \boldsymbol{u}\cdot\boldsymbol{R} + \frac12D^2(\nabla\cdot\boldsymbol{u})^2 + D(\nabla\cdot\boldsymbol{u})(\boldsymbol{u}\cdot\nabla H) + \frac12(\boldsymbol{u}\cdot\nabla H)^2 \right)\cdot d\boldsymbol{x} 
\\
&\qquad - \left(A\exp\frac{2s}{D}+ g\right)\nabla(D-H) - \frac{D}{2}\nabla\left( A\exp\frac{2s}{D}\right)\cdot d\boldsymbol{x}\,.
\end{aligned}
\end{equation}
Equating the left and right hand sides of the equations yields
\begin{equation}\label{eqn:u_GN}
    \p_t\boldsymbol{u} + \boldsymbol{u}\cdot\nabla \boldsymbol{u} + f\boldsymbol{u}^\perp  = \frac{g}{2}D\nabla b + gb\nabla (D-H) + \frac{1}{D} \nabla \Big(D^2 \frac{dF}{dt}\Big) - \frac{dG}{dt} \nabla H \,,
\end{equation}
where $f=\nabla^\perp\cdot\boldsymbol{R}$. Compared to the moist shallow water model, the additional non-hydrostatic terms do not influence the variational derivatives with respect to the entropy, and the sum of the variational derivatives with respect to the depth contributions is unchanged. Thus, the model is closed by the same equations \eqref{eqn:depth_SW} and \eqref{eqn:buoyancy_SW}. 

The thermodynamics is unchanged by the presence of nonhydrostatic terms since they do not influence the variation of the Lagrangian in $s$. Thus, as for the moist shallow water model, the system is augmented with the buoyancy equation \eqref{eqn:buoyancy_SW}, repeated here as
\begin{equation}
    \partial_t b + u\cdot\nabla b 
    =\frac{2jL}{gD\left(-D+2H - \frac{2B}{A}\right)} \,.
\label{eqn:buoyancy_SW1}\end{equation}

As in the previous section on the moist shallow water model, we may write a Kelvin-Noether theorem for these equations
\begin{equation}
\label{eq:Kelvin_MGN}
\frac{d}{dt}
\oint_{c(u)} \left( \boldsymbol{u} + \boldsymbol{R} - \frac{1}{D}\nabla(D^2F) + G\nabla H \right)\cdot d\boldsymbol{x}   = \, \frac{g}{2} \oint_{c(u)}  b\nabla (D-2H)  \cdot d\boldsymbol{x}
\,.
\end{equation}
Likewise, the potential vorticity formulation for this model is
\begin{equation}
    \p_tq_{GN} + \boldsymbol{u}\cdot\nabla q_{GN} = \frac{g}{2D} \nabla^\perp b\cdot \nabla(D-2H) \,,
\end{equation}
where the Green-Naghdi potential vorticity is defined as
\begin{equation}
    q_{GN} = \frac{1}{D}\left( f + \nabla^\perp\cdot\boldsymbol{u}+ \nabla^\perp\cdot\left(- \frac{1}{D}\nabla(D^2F) + G\nabla H \right) \right) \,,
\end{equation}
and features the additional terms arising in the momentum mapping for Green-Naghdi dynamics. As for the shallow water case, and following the discussion of the general case in Appendix \ref{app: C-EnergyConserved}, the moist Green-Naghdi model conserves an energy.
\begin{proposition}
    The moist Green-Naghdi model, given by equations \eqref{eqn:depth_SW}, \eqref{eqn:u_GN}, and \eqref{eqn:buoyancy_SW}, conserves the total energy
    \begin{equation}
    \begin{aligned}
        E_{MGN} &= \int_\Omega \frac{D|u|^2}{2} + \frac16 D^3(\nabla\cdot\boldsymbol{u})^2 + \frac12D^2 (\nabla\cdot\boldsymbol{u})\left(\boldsymbol{u}\cdot\nabla H\right)  + \frac12D\left(\boldsymbol{u}\cdot\nabla H\right)^2 
        \\
        &\qquad\qquad + \frac12 gbD(D-2H) + DB\exp\frac{2s}{D} - D\frac{p_{00}}{\rho_0} - LD_v\,d^2x \,.
    \end{aligned}
    \end{equation}
\end{proposition}

\section{Summary and Outlook}\label{sec: summary}

Two dimensional shallow water models are useful both for studying processes in an idealised setting and for studying the behaviour of numerical discretisations efficiently before looking into 3D models. By following the geometric framework of \cite{gay2019variational} in this article, we have demonstrated the variational derivations of moist hydrodynamics in the thermal hydrostatic shallow water and thermal non-hydrostatic Green-Naghdi models. These derivations in the enthalpic thermodynamic representation provide models with energy conservation and the Kelvin circulation theorem. Both of these properties emerge naturally from the variational framework. The energy conserving formulation is particularly important for long-time simulations in which energy transfer plays a significant role.

The main deviation
from previous models of this type is the latent heat release term in the buoyancy equation \eqref{eqn: b-dynamics}, 
\begin{equation}
\frac{2jL}{gD\left(-D+2H - \frac{2B}{A}\right)}
\,.\end{equation}
This term has an additional factor necessary for energy conservation, compared to the latent heat release of \cite{zerroukat2015moist},
for example. Also, the latent heat $L$ does not appear explicitly in the final thermal shallow water motion equation \eqref{eqn: motion} with moisture dynamics. Instead, the latent heat influences the fluid motion implicitly through its effect on the evolution of the buoyancy in equation \eqref{eqn: b-dynamics}.

The results for the non-hydrostatic thermal Green-Naghdi equation with moisture demonstrates the generality of the variational approach within the class of moist thermal shallow water models, since the hydrostatic and non-hydrostatic models share the same latent heat release term in their respective buoyancy evolution equations.

In future work, we plan to use the variational approach to investigate structure-preserving discretisations for computational simulations and make comparisons with other moisture parameterisations. In particular, we are interested in deriving energy-conserving time discretisations, and using them to examine the dynamics of entropy 
\eqref{eq:entropy repeated} in that setting.

\paragraph{Acknowledgements} The authors are grateful to Thomas Bendall, Christopher Eldred, Fran\c{c}ois Gay-Balmaz, and John Thuburn for useful discussions about this work. Oliver Street acknowledges funding for a research fellowship from Quadrature Climate Foundation, which has supported his contribution to this project. Darryl Holm was partially supported during the present work by European Research Council (ERC) Synergy grant “Stochastic Transport in Upper Ocean Dynamics” (STUOD) – DLV-856408.


\bibliographystyle{apalike}
\bibliography{moisture.bib}

\appendix

\section{Parameterising moisture in Hamilton's Principle}\label{app:formulation}

This Appendix provides a special case of the formulation of \cite{gay2019variational}, in the case
without diffusive fluxes. As described in the main text, we have species $k\in \mathcal{K}$, with $D=\sum_k D_k$,
with species layer depth contributions evolving according to
\begin{equation}
\bar{\mathscr{D}}_t D_k = j_k,
\end{equation}
using the abbreviated notation
\begin{equation}
\mathscr{D}_t = \pp{}{t} + \boldsymbol{u}\cdot \nabla, \quad
\bar{\mathscr{D}}_t = \pp{}{t} + \nabla\cdot(\boldsymbol{u}\cdot) \,.
\end{equation}
We shall also write
\begin{equation}
\mathscr{D}_\delta = \delta + \boldsymbol{v}\cdot \nabla \,, \quad
\bar{\mathscr{D}}_\delta = \delta + \nabla\cdot(\boldsymbol{v}\cdot) \,.
\end{equation}
We will be working in the two dimensional setting relevant to the body of the article, demoting our two dimensional spatial domain by $\Omega \subseteq \mathbb{R}^2$.

\begin{remark}[Notation]\label{rmk:notation}
    As in the main text, we will denote vector fields and their dual elements using a bold font when the basis is not included. That is, vector fields are represented by $u=\boldsymbol{u}\cdot\nabla$ and their dual elements ($1$-form densities) as $\delta\ell / \delta u = \delta\ell/\delta\boldsymbol{u} \cdot d\boldsymbol{x}\otimes d^2x$, where $d\boldsymbol{x}$ denotes the dual basis to $\nabla$. Furthermore, our densities are denoted by $D_kd^2x$, such that $D_k$ represents the function valued part of this only.

    Duality pairings are denoted by angular brackets $\scp{\,\cdot\,}{\,\cdot\,}$ and, as such, variational derivatives are defined as in equations \eqref{eqn:var_derivs_def_2D} and \eqref{eqn:var_derivs_def_3D}.
\end{remark}

Given a reduced Lagrangian $\ell[u,s,\{D_k\}_{k\in \mathcal{K}}]$ depending 
on velocity $u$ and entropy $s$ as well as the layer depth contributions $\{D_k\}_{k\in \mathcal{K}}$
we consider the constrained variational principle
\begin{equation}
\begin{aligned}
0 &= \delta \int_0^T \ell[u, \{D_k\}, s] + \langle\underbrace{ (s-\sigma)d^2x, \mathscr{D}_t\gamma\rangle
+ \sum_k \langle D_kd^2x, \mathscr{D}_t w_k }_{\hbox{Clebsch constraints}}\rangle dt 
\\
&= \delta \int_0^T \left(\ell[u, \{D_k\}, s] + \int_\Omega (s-\sigma)\mathscr{D}_t\gamma \,d^2x + \sum_k \int_\Omega D_k\mathscr{D}_tw_k\,d^2x \right) \,dt
\,,
\end{aligned}
\end{equation}
with constraints,
\begin{align}
\delta u & = \p_t{\xi} + [u,\xi] 
= \big(\p_t{\boldsymbol{\xi}} + (\boldsymbol{u}\cdot \nabla) \boldsymbol{\xi} 
- (\boldsymbol{\xi}\cdot \nabla) \boldsymbol{u} \big)\cdot\nabla 
\label{eqn:constrained_vars}
\\
\nonumber
&=: (\p_t\boldsymbol{\xi} + [\boldsymbol{u},\boldsymbol{\xi}])\cdot\nabla\,, \\
\label{eq:sigma}
\dede{\ell}{s}\big(\p_t{\sigma} + \nabla\cdot(\sigma \boldsymbol{u})\big) & = \dede{\ell}{s}\bar{\mathscr{D}}_t\sigma  = \sum_k j_k\mathscr{D}_tw_k \,, \\
\dede{\ell}{s}\big(\delta \sigma + \nabla\cdot(\sigma \boldsymbol{\xi})\big) & = \dede{\ell}{s}\bar{\mathscr{D}}_\delta \sigma  = \sum_k j_k \mathscr{D}_\delta w_k \,.
\end{align}
The variations in $s$, $\{D_k\}$ and $\gamma$ are free. Here, $s$ plays the role of entropy density, $\sigma$ is an auxiliary density variable whose convective derivative $\bar{\mathscr{D}}_t\sigma$ turns out to be the entropy generation rate, and $\gamma$ and $\{w_k\}_{k\in \mathcal{K}}$ are Lagrange multipliers enforcing the entropy generation and conservation of species layer
depth contributions respectively.

Proceeding, we find
\begin{align}
\begin{split}
0  = & \int_0^T \int_\Omega  \underbrace{\left(\dede{\ell}{\boldsymbol{u}} + (s-\sigma)\nabla \gamma
+ \sum_k D_k \nabla w_k\right)}_{\hbox{Total momentum}}\cdot\left( \dot{\boldsymbol{\xi}} + [\boldsymbol{u},\boldsymbol{\xi}] \right) \,d^2x \\
&\quad + \sum_k \int_\Omega \left ( \dede{\ell}{D_k} + \mathscr{D}_tw_k \right) \delta D_k \,d^2x
+ \sum_k \int_\Omega D_k  \mathscr{D}_t\delta w_k \,d^2x \\
&\quad + \int_\Omega\delta s \left(\dede{\ell}{s} + \mathscr{D}_t\gamma \right)d^2x 
- \int_\Omega \delta \sigma \mathscr{D}_t\gamma \,d^2x
+ ( s-\sigma) \mathscr{D}_t\delta \gamma \,d^2x \,dt \,.
\end{split}
\label{eq:VP}
\end{align}
Stationary of the action with respect to variations $\delta s$ implies the relation $\mathscr{D}_t\gamma=-\dede{\ell}{s}$. Then, from the relations in \eqref{eq:VP} one finds
\begin{align}
-\int_\Omega \delta \sigma \mathscr{D}_t \gamma \,d^2x & = \int_\Omega  \dede{\ell}{s}\delta\sigma \,d^2x \\
& = \int_\Omega \sum_k j_k \mathscr{D}_\delta w_k - \dede{\ell}{s}\nabla\cdot(\sigma \boldsymbol{\xi}) \,d^2x \,.
\end{align}
Now we substitute back into \eqref{eq:VP} to get
\begin{equation}
\begin{aligned}
0 = & \int_0^T 
\int_\Omega  \left(\dede{\ell}{\boldsymbol{u}} + (s-\sigma)\nabla \gamma
+ \sum_k D_k \nabla w_k\right)\cdot\left( \dot{\boldsymbol{\xi}} + [\boldsymbol{u},\boldsymbol{\xi}] \right) \,d^2x 
\\
&\quad + \int_\Omega \sum_k j_k \mathscr{D}_\delta w_k\,d^2x - \int_\Omega \dede{\ell}{s}\nabla\cdot(\sigma \boldsymbol{\xi}) \,d^2x
\\
&\quad + \sum_k \int_\Omega \left ( \dede{\ell}{D_k} + \mathscr{D}_tw_k \right) \delta D_k \,d^2x + \sum_k \int_\Omega D_k  \mathscr{D}_t\delta w_k \,d^2x + ( s-\sigma) \mathscr{D}_t\delta \gamma \,d^2x dt \,.
\end{aligned}
\end{equation}
Integrating by parts in time, and
using the time boundary conditions for variations, we have
\begin{align}
0 = & \int_\Omega \left( -\pp{}{t}\dede{\hat{\ell}}{\boldsymbol{u}}\cdot\boldsymbol{\xi} + 
\dede{\hat{\ell}}{\boldsymbol{u}}\cdot 
[\boldsymbol{u}, \boldsymbol{\xi}] 
+ \sum_k j_k \boldsymbol{\xi}\cdot \nabla w_k  - \dede{\ell}{s}\nabla\cdot(\sigma\boldsymbol{\xi}) \right)\,d^2x \,,
\quad \hbox{for all } \boldsymbol{\xi} 
\label{eqn:varxi}
\,,\\
 0 & =  \sum_k \int_\Omega \left ( \dede{\ell}{D_k} + \mathscr{D}_tw_k \right) \delta D_k \,d^2x \,, \quad \hbox{for all } \delta D_k 
 \label{eqn:varD}
 \,,\\
0  = & \int_\Omega \left(-\pp{D_k}{t} \delta w_k 
+  D_k \boldsymbol{u}\cdot\nabla \delta w_k 
+  j_k \delta w_k \right)\,d^2x \,,\quad \hbox{for all } \delta w_k
\label{eqn:varw}
\,,\\
0 = & \int_\Omega \left(-\pp{(s-\sigma)}{t} \delta \gamma  + (s-\sigma) \boldsymbol{u}\cdot\nabla\delta\gamma \right)\,d^2x \,,\quad \hbox{for all } \delta\gamma
\,,
\label{eqn:vargamma}
\end{align}
where $\dede{\hat{\ell}}{\boldsymbol{u}}$ is the total momentum, appearing in equation \eqref{eqn:varxi} as
\begin{equation}
\dede{\hat{\ell}}{\boldsymbol{u}} = \dede{\ell}{\boldsymbol{u}} + (s-\sigma)\nabla \gamma
+ \sum_k D_k \nabla w_k \,.
\end{equation}
To derive the equations of motion, it remains to integrate by parts in space to remove derivatives from the arbitrary variables. This operation naturally yields the coordinate expressions of the Lie derivative and hence, to simplify our forthcoming derivations, we will give the resulting equation in both coordinate and coordinate-free geometric notations.
\begin{remark}[Integrating by parts in space and the Lie derivative]
    Integrating by parts in space is required to produce various terms in the momentum equation. We will assume natural boundary conditions which ensure that the boundary contributions vanish. A forcing term on the right hand side of our momentum equation is produced by
    \begin{equation*}
        \int_\Omega - \dede{\ell}{s}\nabla\cdot(\sigma\boldsymbol{\xi)}\,d^2x = \int_\Omega \sigma\boldsymbol{\xi}\cdot\nabla\dede{\ell}{s}\,d^2x = \scp{\xi}{d\dede{\ell}{s}\otimes(\sigma d^2x)}\,,
    \end{equation*}
    where, in the final equality, we have used the duality pairing, exterior derivative $d$, tensor product, $\otimes$, and (as in \cite{gay2019variational}) interpreted our variable $\sigma d^2x$ as a density. The need for the additional geometric notation comes from the fact that the dual space to the space of vector fields is the space of $1$-form densities.

    In equations \eqref{eqn:varw} and \eqref{eqn:vargamma}, we produce terms of the form
    \begin{equation*}
        \int_\Omega D_k\boldsymbol{u}\cdot\nabla\delta w_k \,d^2x = \int_\Omega - \nabla\cdot(D_k\boldsymbol{u})\delta w_k\,d^2x =: \scp{-\mathcal{L}_u(D_kd^2x)}{\delta w_k} \,.
    \end{equation*}
    Denoting the $i$-th component of $\delta\hat\ell / \delta\boldsymbol{u}$ by $\delta\hat\ell / \delta u^i$, we have that
    \begin{equation*}
    \begin{aligned}
        \int_\Omega - \dede{\hat{\ell}}{\boldsymbol{u}}\cdot [\boldsymbol{u}, \boldsymbol{\xi}] \,d^2x &= \int_\Omega \dede{\hat{\ell}}{\boldsymbol{u}}\cdot \left( - \boldsymbol{u}\cdot\nabla\boldsymbol{\xi} + \boldsymbol{\xi}\cdot\nabla\boldsymbol{u} \right) =  \int_\Omega - \dede{\hat\ell}{u^i} u^j\pp{\xi^i}{x^j} \,d^2x + \int_\Omega \dede{\hat\ell}{u^j}\xi_i\pp{u^j}{x^i} \,d^2x
        \\
        &= \int_\Omega \xi^i\left( \pp{}{x^j}\left( \dede{\hat\ell}{u^i} u^j \right) + \dede{\hat\ell}{u^j}\pp{u^j}{x^i} \right)\,d^2x 
        \\
        &= \int_\Omega \boldsymbol{\xi}\cdot\left( \boldsymbol{u}\cdot\nabla \dede{\hat\ell}{\boldsymbol{u}} + \dede{\hat\ell}{u^j}\nabla u^j + \dede{\hat\ell}{\boldsymbol{u}}(\nabla\cdot\boldsymbol{u}) \right)\,d^2x 
        \\
        &= \int_\Omega \boldsymbol{\xi}\cdot\left( - \boldsymbol{u}\times {\rm curl}\dede{\hat\ell}{\boldsymbol{u}} + \nabla\left(\boldsymbol{u}\cdot \dede{\hat\ell}{\boldsymbol{u}}\right) + \dede{\hat\ell}{\boldsymbol{u}}(\nabla\cdot\boldsymbol{u}) \right)\,d^2x = \scp{\mathcal{L}_u\dede{\hat\ell}{u}}{\xi} \,,
    \end{aligned}
    \end{equation*}
    where we have given two equivalent coordinate forms of the Lie derivative and provided the coordinate free description in terms of the duality pairing. In practice, our variational derivative with respect to $u = u^i \p/\p x^i$ will produce the tensor product of a $1$-form $\alpha = \alpha_idx^i$ with a density of the form $Dd^2x$. The Lie derivative computed above is then
    \begin{equation*}
        \mathcal{L}_u(D\alpha\otimes d^2x) = D\mathcal{L}_u\alpha \otimes d^2x + \alpha \otimes \mathcal{L}_u(Dd^2x) = \left(D\mathcal{L}_u\alpha + \alpha \nabla\cdot(D\boldsymbol{u}) \right)\otimes d^2x \,,
    \end{equation*}
    where $\mathcal{L}_u\alpha$ is of the form computed above.
\end{remark}
We thus have the language to give our motion equations for the total momentum and three auxiliary equations, in a coordinate-free geometric form, as
\begin{align}
\left(\pp{}{t}+\mathcal{L}_u\right)&\left(\dede{\ell}{u}
+ d\gamma\otimes (s-\sigma)d^2x +\sum_k dw_k \otimes D_kd^2x\right) =
\sum_k d w_k \otimes j_kd^2x + d\dede{\ell}{s}\otimes \sigma\,d^2x 
\label{eq:wk}
\,,\\
\left(\pp{}{t}+\mathcal{L}_u\right)& w_k  = -\dede{\ell}{D_k}, \quad  
\left(\pp{}{t}+\mathcal{L}_u\right)(D_kd^2x)  = j_kd^2x \,, \quad
\left(\pp{}{t}+\mathcal{L}_u\right)(s-\sigma)d^2x  = 0 \,. \label{eq:s-sigma}
\end{align}
We see that, in this notation, $(\p_t + \mathcal{L}_u)$ denotes either $\mathscr{D}_t$ or $\bar{\mathscr{D}}_t$ depending on whether it acts on scalars or densities. Substituting the each of the three equations in \eqref{eq:s-sigma} into equation \eqref{eq:wk}, and using $\mathscr{D}_t\gamma=-\delta\ell/\delta s$, then gives the simplified form
\begin{equation}
\label{eq:eliminated EP}
\left(\pp{}{t} + \mathcal{L}_u\right)\dede{\ell}{u}  = \sum_k d\dede{\ell}{D_k}
\otimes D_kd^2x + d\dede{\ell}{s}\otimes sd^2x \,,
\end{equation}
or, in coordinates, we have the system
\begin{align}
    \pp{}{t}\dede{\ell}{\boldsymbol{u}} + \boldsymbol{u}\cdot\nabla \dede{\ell}{\boldsymbol{u}} + \dede{\ell}{u^j}\nabla u^j + \dede{\ell}{\boldsymbol{u}}(\nabla\cdot\boldsymbol{u}) &= \sum_k D_k\nabla\dede{\ell}{D_k} + s \nabla\dede{\ell}{s}
    \,,\\
    \mathscr{D}_tw_k = -\dede{\ell}{D_k} \,,\quad \bar{\mathscr{D}}_tD_k = j_k \,,\quad \bar{\mathscr{D}}_t(s-\sigma) = 0 \,.
\end{align}
Next, applying $(\p_t + \mathcal{L}_u)(Dd^2x)=0$ for $D=\sum_k D_k$ allows one to write the previous momentum equation in `Kelvin--Noether' circulation form, as 
\begin{equation}
\label{eq:Kelvin EP form}
\left( \pp{}{t} + \mathcal{L}_u \right)
\left(\frac{1}{D}\dede{\ell}{u}\right)  = \left(\sum_k \frac{D_k}{D} d\dede{\ell}{D_k}
 + \frac{s}{D}d\dede{\ell}{s} \right) \,,
\end{equation}
where we are denoting the division of a $1$-form density by a density as an operation which produces a $1$-form as: 
\begin{equation*}
    \frac{1}{D}\dede{\ell}{u} = \frac{1}{D}\dede{\ell}{\boldsymbol{u}}\cdot d\boldsymbol{x} \,.
\end{equation*}
In coordinates, this form of the momentum equation is
\begin{equation}
    \pp{}{t}\frac1D\dede{\ell}{\boldsymbol{u}} + \boldsymbol{u}\cdot\nabla \frac1D\dede{\ell}{\boldsymbol{u}} + \frac1D\dede{\ell}{u^j}\nabla u^j + \frac1D\dede{\ell}{\boldsymbol{u}}(\nabla\cdot\boldsymbol{u}) = \sum_k \frac{D_k}{D}\nabla\dede{\ell}{D_k} + \frac{s}{D} \nabla\dede{\ell}{s} \,.
\end{equation}
This equation directly implies Kelvin's circulation theorem \citep{HMR1998}, in the form
\begin{equation}
\label{eq:Kelvin EP form}
\frac{d}{dt}
\oint_{c(u)}\frac{1}{D}\dede{\ell}{u}  = \oint_{c(u)}\left(\sum_k \frac{D_k}{D} d\dede{\ell}{D_k}
 + \frac{s}{D}d\dede{\ell}{s} \right) \,.
\end{equation}
At this point, we have eliminated the terms $\{w_k\}$ and $\gamma$ in \eqref{eq:VP}. 
To eliminate $\sigma$ as well, we recall the second constraint in
\eqref{eq:sigma}, which can be combined with \eqref{eq:s-sigma} to give
\begin{equation}
\label{eq:entropy}
\left(\pp{}{t}+ \mathcal{L}_u\right)(sd^2x) = -\frac{1}{\dede{\ell}{s}}\sum_k j_k\dede{\ell}{D_k}d^2x \,,\quad\hbox{or in coordinates as}\quad {\mathscr{D}}_ts = - \frac{1}{\dede{\ell}{s}}\sum_k j_k\dede{\ell}{D_k} \,.
\end{equation}

\paragraph{Potential vorticity}
The Kelvin-Noether form of the momentum equation implies a potential vorticity formulation, which is found by taking an exterior derivative on each side of the equation as
\begin{equation}
    \left(\pp{}{t}+\mathcal{L}_u\right)d\left(\frac{1}{D}\dede{\ell}{u}\right) = d\left(\sum_k \frac{D_k}{D} d\dede{\ell}{D_k}
 + \frac{s}{D}D\dede{\ell}{s} \right) \,.
\end{equation}
Since $\frac{1}{D}\dede{\ell}{u}$ is a 1-form on a two dimensional domain, its exterior derivative (the potential vorticity) corresponds to a dot product with the skew gradient $\nabla^\perp = (-\p_y,\p_x)$. Moreover, since $\rho_k$ and $s$ are densities, the variational derivatives of the Lagrangian with respect to these quantities is a function whose exterior derivative corresponds to a gradient. Thus, since the potential vorticity is a function, the potential vorticity equation in vector calculus notation reads
\begin{equation}\label{eqn:general_PV}
    \p_tq+\boldsymbol{u}\cdot\nabla q = \nabla^\perp\cdot\left(\sum_k \frac{D_k}{D} \nabla\dede{\ell}{D_k}
 + \frac{s}{D}\nabla\dede{\ell}{s} \right) \,,
\end{equation}
where
\begin{equation}
    q = \nabla^\perp \cdot \left( \frac{1}{D}\frac{\delta\ell}{\delta\boldsymbol{u}} \right) \,.
\end{equation}

\section{Enthalpic Lagrangian for 3D thermal Boussinesq equations}\label{app:Lagrangian}
In this appendix, we first review the Euler--Poincar\'e derivation, as it is given in \cite{eldred2021thermodynamically}, of the three-dimensional thermal Boussinesq equation via variations of the thermodynamic enthalpy of the medium, which in passing represents the buoyancy in terms of pressure and entropy. Following this, we discuss the definition of buoyancy required for deducing our moist shallow water equations. We begin with the Lagrangian for the 3D inhomogeneous rotating Euler's equation, a standard Lagrangian for geophysical flows, as it appears in \cite{eldred2021thermodynamically}. This is given by
    \begin{equation}
    \label{eqn:Euler_Lagrangian}
        \ell_{E} = \int \frac{\rho|\boldsymbol{v}|^2}{2} + \rho \boldsymbol{v}\cdot\boldsymbol{R} - \rho h(p,\eta) + p - \rho\phi \,d^3x \,,
    \end{equation}
where $v$ is fluid velocity, $\rho$ is the advected mass density, $p$ is pressure, $\eta$ is entropy, $\phi$ is the geopotential, and $h$ is scalar enthalpy. Here, $\boldsymbol{R}$ is defines the rotational effect and is such that ${\rm curl}\boldsymbol{R} = f\boldsymbol{\hat{z}}$, where $f$ is the Coriolis parameter. Notice that the Lagrangian here is expressed in terms of specific enthalpy, $h(p,\eta)$, related to specific internal energy $e(\rho,\eta)$ by $h(p,\eta) = e(\rho,\eta) + p/\rho$. This description offers a convenient passage to the Euler-Boussinesq model. The arbitrary variation of the Lagrangian in $p$ defines the density in terms of enthalpy. The constrained Euler-Poincar\'e variations with respect to $u$, $\rho$, and $\eta$, together with the 1st law of thermodynamics $dh = (1/\rho) dp + Td\eta$, imply Euler's equation. See \cite{HMR1998} for further details about this type of symmetry-broken variational principle for compressible flows.

Following \cite{eldred2021thermodynamically}, we linearise the Lagrangian around a background pressure profile $p_0$ to write $p=p_0+p'$ and find that the expansion to first order in $p'$ produces a Lagrangian for pseudoincompressible models \citep{cotter2014variational}. This expansion happens in the potential energy term, as
\begin{equation}\label{eqn:pe_expansion}
\begin{aligned}
    \rho gz + \rho e(\rho,\eta) &= \rho gz + \rho h(p,\eta) - p
    \\
    &\approx \rho gz + \rho h(p_0,\eta) + \rho p' \frac{\partial h(\tilde p , \eta)}{\partial \tilde p}\bigg|_{\tilde p = p_0} - p_0 - p' 
    \\
    &= \rho gz + \rho h(p_0,\eta)-p_0 - p'\left( 1 - \frac{\rho}{\rho(p_0,\eta)} \right) \,.
\end{aligned}
\end{equation}
The expansion of enthalpy produces a term $(\partial h / \partial p)(p_0,\eta)=1/\rho(p_0,\eta)$ and if this is taken to be a constant $1/\rho_0$, then we have the Boussinesq approximation\footnote{If, instead, this term is taken to be the reciprocal of the stratification mass density profile $1/\rho_0(x)$, then we obtain anelastic models.}. The resulting Lagrangian then corresponds to an \emph{incomressible} Boussinesq model, with incompressibility enforced by $p'$ acting as a Lagrange multiplier. Namely, our Lagrangian is
\begin{equation}\label{eqn:EB_Lagrangian}
    \ell_{EB}(\boldsymbol{v},\rho,\eta;p') = \int \frac{\rho|\boldsymbol{v}|^2}{2} + \rho \boldsymbol{v}\cdot\boldsymbol{R} - \rho\phi - \rho h(p_0,\eta) + p'\left( 1-\frac{\rho}{\rho_0} \right) + p_0\,d^3x \,,
\end{equation}
where $\rho_0$ is a constant and $\phi$ is the geopotential. The variational derivatives of this Lagrangian are computed as
\begin{align}
    \frac{\delta\ell_{EB}}{\delta v} &= \rho(v+R)
    \,,\\
    \frac{\delta\ell_{EB}}{\delta \rho} &= \frac12|\boldsymbol{v}|^2 + \boldsymbol{v}\cdot\boldsymbol{R} - \phi - h(p_0,\eta) - \frac{p'}{\rho_0}
    \,,\\
    \frac{\delta\ell_{EB}}{\delta \eta} &= - \rho \frac{\partial h}{\partial\eta} 
    \,,\\
    \delta p' &:\quad \rho = \rho_0
    \,,
\end{align}
where we have defined the variations with respect to the $L^2$ pairing as 
\begin{equation}\label{eqn:var_derivs_def_3D}
\begin{aligned}
    \delta \ell(v(\boldsymbol{x},t)) &:= \frac{\p}{\p\epsilon}\bigg|_{\epsilon=0} \ell(v(\boldsymbol{x},t,\epsilon)) = \scp{\frac{\delta\ell}{\delta v}} {\frac{\p v}{\p\epsilon}\bigg|_{\epsilon=0}} = \scp{\frac{\p\ell}{\p v}}{ \delta v } := \int\sum_{j=1}^3\frac{\p\ell}{\p v^j} \,\delta v^j d^3x = \int \frac{\delta\ell}{\delta\boldsymbol{v}}\cdot\delta\boldsymbol{v}\,d^3x
    \,,\\
    \delta\ell(\eta(\boldsymbol{x},t)) &= \left\langle \frac{\delta\ell}{\delta \eta}\,d^3x , \delta\eta  \right\rangle = \int \frac{\delta\ell}{\delta\eta}\delta \eta \,d^3x
    \,,\quad\hbox{and}\quad
    \delta\ell(\rho(\boldsymbol{x},t)) = \left\langle \frac{\delta\ell}{\delta\rho} , \rho \,d^3x \right\rangle = \int \frac{\delta\ell}{\delta\rho}\delta \rho \,d^3x
    \,.
\end{aligned}
\end{equation}
Note that, unlike much of the literature on geometric modelling of continua, for simplicity we have defined the variational derivatives such that they do not include their basis as a differential form (or vector field) type object. Instead, they are vectors or scalar functions. To apply Hamilton's Principle to the corresponding action, we make use of standard semidirect produce Euler-Poincar\'e theory \cite{HMR1998}. That is, the variation of $\boldsymbol{v}$ is the standard constrained variation of a velocity vector field, as in equation \eqref{eqn:constrained_vars}. The variation of $\rho$ and $\eta$ are constrained as $\delta\rho = -{\rm div}(\rho\boldsymbol{\xi})$, $\delta\eta = -\boldsymbol{\xi}\cdot\nabla\eta$, due to the assumption that the variables are advected as a density and scalar respectively. The Euler-Poincar\'e momentum equation is then
\begin{equation}
\begin{aligned}
    (\p_t + \mathcal{L}_v)\left( \rho (\boldsymbol{v}+\boldsymbol{R})\cdot d\boldsymbol{x}\otimes d^3x \right) 
    &= \rho\, d\!\left( \frac12|\boldsymbol{v}|^2 + \boldsymbol{v}\cdot\boldsymbol{R} - \phi - h(p_0,\eta) - \frac{p'}{\rho_0} \right) \otimes d^3x  + \rho(p_0,\eta) \frac{\partial h}{\partial\eta} d\eta \otimes d^3x 
    \,,\\
    (\p_t + \mathcal{L}_v)\eta &= 0
    \,,\quad
    (\p_t + \mathcal{L}_v)(\rho d^3x) = 0
    \,.
\end{aligned}
\label{eqn: SW-EP}
\end{equation}
Note that this equation set can also be found by varying in entropy density $s=\rho\eta$. Stationarity of the action in $p'$ implies $\rho=\rho_0$, which implies that $\mathrm{div}\boldsymbol{v}=0$. Then, following \cite{eldred2021thermodynamically}, we define \emph{buoyancy} by 
\begin{equation}
\label{eqn: buoyancy defn}
b(\phi,\eta) := -\frac{\partial}{\partial\phi} \left( h(p_0(\phi),\eta) + \phi \right) \,,
\end{equation}
where we have fixed $\eta$ and taken a derivative in the geopotential. Since this relation gives a definition of buoyancy in terms of advected variables, we find that $b$ is also advected. The forcing terms depending on enthalpy and geopotential can be expressed as
\begin{equation}
    \rho d \left( -h - \phi \right) + \rho \frac{\partial h}{\partial \eta}d\eta = -\rho \frac{\partial}{\partial \phi}\left( h+\phi\right)d\phi - \rho \frac{\partial h}{\partial \eta}d\eta + \rho \frac{\partial h}{\partial \eta}d\eta =  \rho b\, d \phi \,.
\end{equation}
Thus, the Euler-Poincar\'e momentum equation in \eqref{eqn: SW-EP} finally simplifies to the Kelvin--Noether form
\begin{equation}
    (\p_t + \mathcal{L}_v)\big(( \boldsymbol{v} + \boldsymbol{R} )\cdot d\boldsymbol{x} \big)
    = 
    d\left( \frac12|\boldsymbol{v}|^2 + \boldsymbol{v}\cdot\boldsymbol{R} - \frac{p'}{\rho_0} \right) 
    + b\,d\phi \,,
\label{SWKN-form}
\end{equation}
which in turn yields the Kelvin circulation dynamics,
\begin{equation}
\label{eq:Kelvin SW form}
\frac{d}{dt}
\oint_{c(v)}( \boldsymbol{v} + \boldsymbol{R} )\cdot d\boldsymbol{x}  
= \oint_{c(v)}b\,d\phi
\,.
\end{equation}

When expressed in terms of vector calculus notation for standard $x,y,z$ Cartesian coordinate systems, and the geopotential is taken to be $\phi = gz$, one obtains the familiar thermal Euler-Boussinesq equations
\begin{equation}
    \p_t \boldsymbol{v} + \boldsymbol{v}\cdot\nabla\boldsymbol{v} = - \frac{1}{\rho_0}\nabla p' + bg\widehat{z}
    \,,\qquad
    \p_t b + \boldsymbol{v}\cdot\nabla b = 0
    \,,\qquad
    \nabla\cdot\boldsymbol{v} = 0 
    \,.
\end{equation}

\begin{remark}[The definition of buoyancy]
    As discussed by \cite{eldred2021thermodynamically}, the definition of buoyancy in the context of soundproof models is given in equation \eqref{eqn: buoyancy defn} and can be interpreted by using the chain rule
\begin{equation}
\begin{aligned}
    b(\phi,\eta) &= -\frac{\partial h}{\partial p}\bigg|_{p=p_0}\frac{\partial p_0}{\partial\phi} -1 = -\frac{1}{\rho(p_0,\eta)}\frac{\partial p_0}{\partial\phi} - 1
    \\
    &= \frac{\rho_0 - \rho(p_0,\eta)}{\rho(p_0,\eta)} \,,
\end{aligned}
\end{equation}
where, on to move from the first line to the second, we have used the fact that our reference profile is in hydrostatic balance. This definition is consistent with that used by \cite{tailleux2011,tailleux2012}, following earlier contributions by \cite{pauluis2008thermodynamic} and \cite{young2010dynamic} for atmospheric and oceanic flows respectively. Notice that to derive the Lagrangian for the Euler-Boussinesq model, when expanding the enthalpy we needed to set its derivative with respect to pressure to be constant, when evaluated at the hydrostatic reference pressure. If we do the same in the above calculation, we see that the denominator becomes $\rho_0$ and the definition collapses to a more familiar form.
\end{remark}

\section{Energy conservation}\label{app: C-EnergyConserved}

This Appendix provides a proof of energy conservation for variational formulations of the form given in Appendix \ref{app:formulation}. We will be using the geometric notation discussed elsewhere in the other appendices.

We define the energy by Legendre transform, by calling $\hat{\ell}$ the integrand of the action, i.e.
\begin{equation}
\hat{\ell}[u, \gamma, \gamma_t, \{D_k\}, s, \sigma, \{w_k\}, \{\pp{w_k}{t}\}] = \ell 
+ \langle (s-\sigma)d^2x,\mathscr{D}_t\gamma\rangle + \sum_k \langle D_kd^2x, \mathscr{D}_t w_k\rangle \,. 
\end{equation}
Then, 
\begin{equation}
\begin{aligned}
E &= \left\langle \dede{\hat\ell}{u}, u \right\rangle 
+ \left\langle \dede{\hat{\ell}}{\gamma_t} d^2x, \gamma_t\right\rangle 
+ \sum_k \left\langle \dede{\hat{\ell}}{w_{k,t}}d^2x, w_{k,t}\right\rangle
- \hat{\ell}
\\
&= \int_\Omega \dede{\hat\ell}{\boldsymbol{u}}\cdot\boldsymbol{u} + \dede{\hat\ell}{\gamma_t}\gamma_t + \sum_k \dede{\hat\ell}{w_{k,t}}w_{k,t}\,d^2x - \hat\ell \,. 
\end{aligned}
\label{eq:legendre}
\end{equation}
In fact, since $\hat{\ell}$ is affine in $\gamma_t$, $\{w_{k,t}\}$, we can write
\begin{align}
E & = \left\langle \dede{\ell}{u}, u \right\rangle - \ell \,,
\end{align}
however \eqref{eq:legendre} is the starting point for deriving energy conservation \emph{via} a N\"oether-style argument.
We take the time derivative, following the notation introduced in equation \eqref{eqn:var_derivs_def_2D} for duality pairings, summing over $k$ where it appears, and denoting time derivatives by a subscript where appropriate. We have
\begin{equation}
\begin{aligned}
    \dot{E} &= \int_\Omega \left(\pp{}{t}\dede{\hat\ell}{\boldsymbol{u}}\right)\cdot\boldsymbol{u}  + \dede{\hat\ell}{\boldsymbol{u}}\cdot\pp{\boldsymbol{u}}{t} + \left(\pp{}{t}\dede{\hat\ell}{\gamma_t}\right)\gamma_t + \dede{\hat\ell}{\gamma_t}\gamma_{tt} + \sum_k\left[\left(\pp{}{t}\dede{\hat\ell}{w_{k,t}}\right)w_{k,t} + \dede{\hat\ell}{w_{k,t}}w_{k,tt}\right]\,d^2x
    \\
    &\qquad - \int_\Omega \dede{\hat\ell}{\boldsymbol{u}}\cdot \pp{\boldsymbol{u}}{t} + \dede{\hat\ell}{s}s_t + \dede{\hat\ell}{\sigma}\sigma_t + \dede{\hat\ell}{\gamma}\gamma_t + \dede{\hat\ell}{\gamma_t}\gamma_{tt} + \sum_k\left[ \dede{\hat\ell}{D_k}D_{k,t} + \dede{\hat\ell}{w_k}w_{k,t} + \dede{\hat\ell}{w_{k,t}}w_{k,tt}\right]\,d^2x
    \\
    &= \int_\Omega \left(\pp{}{t}\dede{\hat\ell}{\boldsymbol{u}}\right)\cdot\boldsymbol{u} +\left(\pp{}{t}\dede{\hat\ell}{\gamma_t}\right)\gamma_t - \dede{\hat\ell}{s}s_t - \dede{\hat\ell}{\sigma}\sigma_t - \dede{\hat\ell}{\gamma}\gamma_t
    \\
    &\qquad\qquad  + \sum_k\left[ \left(\pp{}{t}\dede{\hat\ell}{w_{k,t}}\right)w_{k,t} - \dede{\hat\ell}{D_k}D_{k,t} - \dede{\hat\ell}{w_k}w_{k,t}  \right]\,d^2x \,.
\end{aligned}
\end{equation}
We look at this statement term by term, noticing that from the definition of $\hat\ell$ and from the equations of motion in Appendix \ref{app:formulation}, we have
\begin{align}
    \int_\Omega \left(\pp{}{t}\dede{\hat\ell}{\boldsymbol{u}}\right)\cdot\boldsymbol{u}\,d^2x &=   \int_\Omega  \dede{\hat{\ell}}{\boldsymbol{u}}\cdot [\boldsymbol{u}, \boldsymbol{u}] + \sum_j j_k \boldsymbol{u}\cdot \nabla w_k  - \dede{\ell}{s}\nabla\cdot(\sigma\boldsymbol{u}) \,d^2x
    \,,\\
    \int_\Omega\left(\pp{}{t}\dede{\hat\ell}{\gamma_t}\right)\gamma_t\,d^2x &= \int_\Omega \pp{(s-\sigma)}{t}\gamma_t\,d^2x
    \,,\\
    \int_\Omega\dede{\hat\ell}{s}s_t\,d^2x &= \int_\Omega \dede{\ell}{s}s_t + (\mathscr{D}_t\gamma) s_t\,d^2x = 0
    \,,\\
    \int_\Omega \dede{\hat\ell}{\sigma}\sigma_t \,d^2x &= \int_\Omega -(\mathscr{D}_t\gamma)\sigma_t\,d^2x = \int_\Omega \dede{\ell}{s}\sigma_t\,d^2x 
    \,,\\
    \int_\Omega \dede{\hat\ell}{\gamma}\gamma_t \,d^2x &= \int_\Omega - \gamma_t\nabla\cdot((s-\sigma)\boldsymbol{u})\,d^2x = 0
    \,,\\
    \int_\Omega \left(\pp{}{t}\dede{\hat\ell}{w_{k,t}}\right)w_{k,t}\,d^2x &= \int_\Omega \pp{D_k}{t}w_{k,t}\,d^2x
    \,,\\
    \int_\Omega \dede{\hat\ell}{D_k}D_{k,t} \,d^2x &= \int_\Omega \dede{\ell}{D_k}D_{k,t} \,d^2x + \int_\Omega D_{k,t}\mathcal{D}_tw_k \,d^2x = 0
    \,,\\
    \int_\Omega \dede{\hat\ell}{w_k}w_{k,t}\,d^2x &= \int_\Omega -\nabla \cdot (D_k\boldsymbol{u})w_{k,t}\,d^2x
    \,.
\end{align}
Thus, we have
\begin{equation}
\begin{aligned}
    \dot{E} &= \int_\Omega  \dede{\hat{\ell}}{\boldsymbol{u}}\cdot \underbrace{[\boldsymbol{u}, \boldsymbol{u}]}_{=0} + \sum_k j_k \boldsymbol{u}\cdot \nabla w_k  - \dede{\ell}{s}\nabla\cdot(\sigma\boldsymbol{u}) \,d^2x
    \\
    &\qquad + \int_\Omega  \underbrace{\left(\pp{(s-\sigma)}{t} + \nabla\cdot((s-\sigma)\boldsymbol{u}) \right)}_{=0} \gamma_t\,d^2x - \underbrace{\int_\Omega \dede{\hat\ell}{s}s_t\,d^2x}_{=0} -\int_\Omega\dede{\ell}{s}\sigma_t\,d^2x
    \\
    &\qquad +\sum_k\int_\Omega \bigg[ \underbrace{\pp{D_k}{t}w_{k,t}  + \nabla \cdot (D_k\boldsymbol{u})w_{k,t}}_{=j_kw_{k,t}} +  \underbrace{\dede{\hat\ell}{D_k}D_{k,t}}_{=0} \bigg]\,d^2x
    \\
    &= \int_\Omega j_k\mathscr{D}_tw_k - \dede{\ell}{s}\bar{\mathscr{D}}_t\sigma  \,d^2x = 0 \,,
\end{aligned}
\end{equation}
where, in the final line, we have used the constraint \eqref{eq:sigma}.

\end{document}